\documentstyle[manuscript,prb,aps]{revtex}
\tightenlines 
\begin{document}
\title{A Unified Algebraic Approach to Few and Many-Body
Correlated Systems}
\author{N. Gurappa\thanks{panisprs@uohyd.ernet.in} and Prasanta K.
Panigrahi\thanks{panisp@uohyd.ernet.in}}
\address{School of Physics,
University of Hyderabad,
Hyderabad,\\ Andhra Pradesh,
500 046 INDIA.}
\maketitle

\begin{abstract}

The present article is an extended version of the paper 
{\it Phys. Rev.} {\bf B 59}, R2490 (1999), where, we have established the
equivalence of the Calogero-Sutherland model to decoupled
oscillators. Here, we first employ the same approach for finding
the eigenstates of a large class of Hamiltonians, dealing with
correlated systems. A number of few and many-body interacting
models are studied and the relationship between their respective
Hilbert spaces, with that of oscillators, is found. This
connection is then used to obtain the spectrum generating algebras
for these systems and make an algebraic statement about correlated
systems. The procedure to generate new solvable interacting models
is outlined. We then point out the inadequacies of the present
technique and make use of a novel method for solving linear
differential equations to diagonalize the Sutherland model and
establish a precise connection between this correlated system's
wave functions, with those of the free particles on a circle. In
the process, we obtain a new expression for the Jack polynomials.
In two dimensions, we analyze the Hamiltonian having Laughlin wave
function as the ground-state and point out the natural emergence of 
the underlying linear $W_{1+\infty}$ symmetry in this approach.

\end{abstract}
\draft

\section{Introduction}

The Calogero-Sutherland model (CSM),\cite{cal1,suther1,suther2}
Sutherland model (SM)\cite{suthd} and their generalizations have
found physical applications in various fields such as the
universal conductance fluctuations in mesoscopic
systems,\cite{ms1,ms2,ms3} quantum Hall
effect,\cite{qhe1,qhe2,qhe3} wave propagation in stratified
fields,\cite{wp} random matrix
theory,\cite{suther1,rmt1,rmt2,rmt3,rmt4,rmt5} chaos,\cite{nala,sim}
fractional statistics,\cite{fs1,fs2,fs3,fs4,fs5}
gravity,\cite{gra1,gra2,gra3} anyons\cite{anyon1,anyon2,anyon3}
and gauge theories.\cite{gt1,gt2} This class of many-body systems
with long-range interactions have been studied quite extensively
in the current literature.\cite{book} Ref.[\onlinecite{moap}] is a
classic review; newer developments have been covered in
Refs.[\onlinecite{book,apo,adh,ajb}]. Interestingly, CSM and SM
type long-range interactions have also manifested in models
dealing with pairing interactions\cite{pi} and phase
transitions.\cite{pt}

The increasing appearance of correlated systems in diverse
physical phenomena makes these exactly solvable models, with
in-built correlation due to the presence of the Jastrow factor in
the eigenfunctions, an ideal testing ground, for ideas ranging
from Haldane statistics\cite{fs2} to black hole
physics.\cite{gra3,gt2,gra4} In recent times, several new
mathematical structures like $S_N$-{\it extended} Heisenberg
algebra,\cite{exop,eop} Yangians\cite{yan1,yan2} and deformed
algebras\cite{dea,vel1,vel2} have been employed for unravelling
the precise structure of the Hilbert space of these systems. The
linearity of the eigen spectra of the CSM and the fact that
interaction only modifies the ground-state has led Calogero to
conjecture the existence of an exact mapping between the CSM and
decoupled oscillators.\cite{cal2} The $S_N$-{\it extended}
Heisenberg algebra, with its oscillator like creation and
annihilation operators, partially achieved the above objective,
although the oscillators were not decoupled and consequently, the
members of the Hilbert space, although linearly independent, were
not orthogonal. Recently, we have been able to show the precise
equivalence of the CSM to decoupled oscillators, without involving
any additional mathematical construct.\cite{prb} The formal
appearance of a new inner product in Ref. [\onlinecite{prb}] has
been explicated recently,\cite{uji1} as also the successful
application of our procedure \cite{sscsm1} to supersymmetric CSM,\cite{sscsm2}
the scattering Hamiltonian\cite{gon} and other related
models.\cite{susy1,susy2,susy3,nnn1,nnn2,nnn3,nnn4} The method
also applies to Hamiltonians, containing other competing
long-range interactions.\cite{bbm} However, the subtleties
involved in extending the present procedure to models based on
root systems other than $A_{N-1}$,\cite{moap} as also to higher
dimensions, still remains to be explored. Furthermore, it can be
seen that the above mentioned method to diagonalize the CSM is
inadequate for the diagonalization of SM and its generalizations.
Unlike the CSM, the eigen spectra of SM is quadratic in nature,
where the interaction modifies the ground state as well as the
excited states. The present article intends to deal with the
various generalizations of the CSM and makes use of a novel method
of solving linear differential equations\cite{cha1,cha2} for
diagonalizing the SM and finding its connection with free
particles on a circle.

The paper is organized as follows.
 Sec. II is devoted to a brief summary of the results obtained
in Ref. [\onlinecite{prb}]. The $B_N$-type CSM and a
recently proposed four particle model in one-dimension\cite{ruhl}
are then studied in detail. In the case of $B_N$ model, it is found
that the underlying spectrum generating algebra is $SU(1,1)$,
unlike the $A_{N-1}$ CSM, where it was the Heisenberg algebra.
This also leads to the straightforward construction of the {\it
linear} $W_{\infty}$ algebra. The eigenfunctions and commuting
constants of motion, behind the quantum integrability, are
presented. We point out in Sec. III, how this method can be used to
construct more general interacting models of the CSM type. Sec. IV
deals with the inadequacies of the procedure used for
diagonalization of the CSM and illustrates a new method of
solving linear differential equations, 
which can be employed for the diagonalization of both CSM and SM.
Through this approach we establish the precise connection between free 
particles on
a circle and interacting particles in SM. In the process, we
obtain a novel expression for the Jack polynomials in terms of the
monomial symmetric functions. In Sec. V, we analyze a
two-dimensional interacting Hamiltonian for electrons under the
influence of both electromagnetic and Chern-Simons gauge
fields.\cite{qhe} This model has Laughlin wave function as the
ground-state and is also relevant for implementing the Jain
picture of the fractional quantum Hall effect.\cite{jain} We
explicitly show the connection of this interacting model with
non-interacting harmonic oscillators, and point out the intricacies involved 
in extending the present procedure to higher dimensions. 
In this process the underlying
linear $W_{1+\infty}$ symmetry algebra with the Laughlin wave
function as the highest weight vector of this algebra is brought out in a
transparent manner. We conclude in the last section after pointing out 
several areas, where the
present procedure can be profitably employed.

\section{Diagonalizing Calogero-Sutherland Type Interacting Model}

In this section, we outline the procedure employed in {\it Phys.
Rev.} {\bf B 59}, R2490 (1999) to diagonalize the CSM and map it 
to a set of free harmonic oscillators.\cite{prb} This equivalence
provides an elegant and straightforward method to construct the complete
set of eigenstates of the CSM, starting from the
symmetrized form of the eigenstates of the harmonic oscillators 
we first employ the same procedure to diagonalize the $B_N$-type Calogero model
and estabhlish the precise connection between the Hilbert spaces of CSM and the
present model. A large
class of CSM type models
in one and higher dimensions can also be solved in an analogous manner.

The mapping of the CSM to decoupled oscillators involves a series
of similarity transformations (ST). The CSM Hamiltonian, in the
units, $\hbar = \omega = m = 1$ (we will use these units
throughout the text, unless specified otherwise) is given by
\begin{equation} \label{csm}
H_{CSM}  =  -{1\over2} \sum_i^N \partial_{x_i}^2 +  {1\over2}  \sum_i^N
x_i^2  +
{g^2\over2} \sum_{{i,j} \atop {i \ne j}}^N {1\over(x_i-x_j)^2} \qquad,
\end{equation}
where, $\partial_{x_i} \equiv \partial/\partial x_i$ and $g^2 > - 1/4$.

The above N-body quantum problem is first solved in
one sector of the configuration space such that $x_1 < x_2 ...< x_N$.
The solutions are then analytically continued to other sectors of the
Hilbert space. This is possible in one dimension, because
the repulsive potential in CSM does not allow the particles to overtake
each other. The many-body system can also be quantized either as bosons or
fermions, without loss of generality. We list the
following definitions for the sake of later convenience:

\begin{eqnarray*}
Z & \equiv & \Pi_{i<j}{(x_i - x_j)}^{\beta} \quad,\\
\hat{A}(\beta) & \equiv & \frac{1}{2}
\sum_i \partial_{x_i}^2 + \beta \sum_{i\ne j}
\frac{1}{(x_i - x_j)}
\partial_{x_i} \quad,\\
\hat E(\beta) & \equiv & \exp\{- \hat A(\beta)/2\} \quad,\\
G & \equiv & \exp\left\{-\frac{1}{2} \sum_i x_i^2\right\} \quad.
\end{eqnarray*}

Combining the above three transformations, we find that the operator,
\begin{eqnarray} \label{that}
\hat T \equiv Z G \hat E(\beta)
\end{eqnarray}
diagonalizes the original $H_{CSM}$ (with $\beta(\beta-1)=g^2)$:
\begin{equation} \label{3}
{\hat T^{-1}} H_{CSM} {\hat T} = \sum_i x_i \partial_{x_i} + E_0 \equiv H_{CSM}^{\prime}\qquad.
\end{equation}
Here $E_0 = N/2 + N(N-1)/2\beta$ is the ground state energy. 
The above equation has also been obtained by Sogo in an entirely different
context.\cite{sogo}

Eq. (\ref{3}) reveals that, starting from a symmetric polynomial basis
one can construct the eigenfunctions of the CSM by an inverse similarity
transformation. The CSM can also be mapped to the decoupled oscillators 
using another inverse similarity transformation:

\begin{equation} \label{vst}
G \hat{E}(0) H_{CSM}^{\prime} \hat{E}^{-1}(0) G^{-1} =
-\frac{1}{2} \sum_i \partial_{x_i}^2 + \frac{1}{2} \sum_i
x_i^2 + (E_0 - \frac{1}{2} N) = H_{{\rm free}}\qquad.
\end{equation}

These results can be succintly written as
\begin{eqnarray}
{\hat O}^{-1} H_{CSM} {\hat O} = H_{\rm {oscillator}} + E_0 - \frac{N}{2} \quad,
\end{eqnarray}
where, $\hat O \equiv \hat T E^{-1}(0) G^{-1}$.
As anticipated by Calogero and indicated by the structure of the
eigenspectrum, the CSM Hamiltonian can be mapped to $N$ free oscillators,
apart from the coupling dependent overall shift of the energy eigenvalues.
Hence, it follows that, the excited energy levels and the degeneracy
structure of both these systems are identical, a fact known since the
original solution of the interacting model. \cite{cal1}

We now proceed to the $B_N$ model and first point out the fact that, 
while applying the various similarity transformations
one needs to ensure that the resulting eigenfunctions are members
of the Hilbert space. This is the origin of differences in the
spectrum generating algebras of various CSM type models. It will
be shown that for the $B_N$ case the underlying algebra is not the
Heisenberg one as in the CSM but the $SU(1,1)$ algebra. The
$B_N$-type CSM (BnCSM) Hamiltonian is given by,\cite{moap}
\begin{equation} \label{hbn}
H_{B_N} = - \frac{1}{2} \sum_{i=1}^N \partial_{x_i}^2
+ \frac {1}{2} \sum_{i=1}^N x_i^2 + \frac {1}{2} g^2
\sum_{{i,j=1}\atop {i\ne j}}^N \left\{\frac {1}{(x_i - x_j)^2} +
\frac {1}{(x_i + x_j)^2}\right\} + \frac{1}{2} g_1^2
\sum_{i=1}^N \frac{1}{x_i^2} \qquad,
\end{equation}
where, $g^2$ and $g_1^2$ are two independent
coupling constants. Since, the ground-state wave function
of $H_{B_N}$, when the system is quantized as bosons, is given by
\begin{equation}
\psi_0 = \prod_{1\le {j < {k\le N}}}|x_i - x_j|^\lambda |x_i + x_j|^\lambda
\prod_k^N |x_k|^{\lambda_1} \exp\{- \frac{1}{2}\sum_i x_i^2\}\qquad,
\end{equation}
one can make the following ST, after confining oneself in one sector of the
configuration space :
\begin{equation} \label{tilbn}
\tilde H \equiv \psi_0^{-1} H \psi_0 = \sum_i x_i \partial_{x_i} +
E_0^\prime + \hat F \qquad.
\end{equation}
Here,
\begin{eqnarray*}
\hat F &\equiv& - \left(\frac{1}{2} \sum_i \partial_{x_i}^2
+ \lambda \sum_{i<j} \frac {1}{(x_i^2 - x_j^2)} (x_i \partial_{x_i} - x_j
\partial_{x_j})  + \lambda_1 \sum_i \frac{1}{x_i}
\partial_{x_i}\right) \quad,\\
g^2 &=& \lambda (\lambda - 1) \quad, \\
g_1^2 &=& \lambda_1 (\lambda_1 - 1) \quad,
\end{eqnarray*}
and the ground-state energy $E_0^\prime$ is given by
$$E_0^\prime = N (\frac{1}{2} + (N-1) \lambda + \lambda_1) \quad.$$
As in the case of the CSM, the eigenfunctions of $\tilde H$ must be
totally symmetric with respect to the exchange of any two particle
coordinates. One can easily establish the following
commutation relation;
\begin{equation} \label{bneo}
[\sum_i x_i \partial_{x_i}\,\,,\,\,\exp\{\hat F/2\}] = - \hat F \exp\{\hat
F/2\}\qquad.
\end{equation}
Making use of Eq. (\ref{bneo}) in Eq. (\ref{tilbn}), one gets
\begin{equation} \label{diabn}
\exp\{- \hat F/2\} \tilde H_{B_N} \exp\{\hat F/2\} = \sum_i x_i
\partial_{x_i} + E_0^\prime \qquad.
\end{equation}
The above two results are identical to the CSM case; this is due the fact
that the Euler operator is only sensitive to the degree of the operator
${\hat F}$.
Hence, the ST by the operator $\hat S \equiv \psi_0
\exp\{\hat F/2\}$ diagonalizes $H_{B_N}$ {\it i.e.},
\begin{eqnarray}
{\hat S}^{-1} H_{B_N} \hat S = \sum_i x_i \partial_{x_i} + E_0^\prime
\quad.
\end{eqnarray}
Furthermore, the following similarity
transformation on Eq. (\ref{diabn}) makes the connection of the BnCSM
with the decoupled oscillators explicit:
\begin{equation}
G \hat E(0) {\hat S}^{-1} H_{B_N} \hat S {\hat E}^{-1}(0) G^{-1} =
-\frac{1}{2} \sum_i \frac{\partial^2}{\partial x_i^2} + \frac{1}{2} \sum_i
x_i^2 + (E_0^\prime - \frac{1}{2} N) \qquad,
\end{equation}
where, $\hat E(0)$ and $G$ are as defined earlier.

\subsection{Mapping between the CSM and BnCSM}

It is interesting to note that, $H_{B_N}$ can also
be made equivalent to the CSM as follows;
\begin{equation}
\hat T \hat S^{-1} H_{B_N} \hat S {\hat T}^{-1} = -
\frac{1}{2} \sum_i \partial_{x_i}^2 + \frac{1}{2} \sum_i x_i^2 +
\sum_{i<j} \frac{\beta (\beta - 1)}{(x_i - x_j)^2} + (E_0^\prime -
E_0)\qquad,
\end{equation}
where, $E_0$ is the ground-state energy of the CSM. One needs to
check the spaces of functions on which all the above ST's are well
defined. This point will be elaborated more after finding out the
Hilbert space of the $B_N$ model. This will also lead to the
precise relationship between the respective eigenspaces of
$H_{CSM}$ and $H_{B_N}$.

\subsection{Hilbert space of BnCSM}

We can make use of Eq. (\ref{diabn}) for the explicit construction of
the eigenfunctions of Eq. (\ref{hbn}). It is easy to
see that an arbitrary homogeneous symmetric function of $x_i$'s 
 \cite{jack4} is an eigenfunction of $\sum_i
x_i \partial_{x_i}$; however, unlike the case of CSM, only the
homogeneous symmetric functions of the
square of the particle co-ordinates are the ones on which the action of
$\exp\{\hat F/2\}$ gives a polynomial solution.  At this moment, it is
also worth noting that
if $\lambda =
1$, then Eq. (\ref{tilbn}) reduces to the differential equation, in $y_i =
x_i^2$
variables for the multivariate Lagurre polynomials.\cite{mvlp} This is the
basic reason why, the
action of the operator $\exp\{\hat F/2\}$ on the homogeneous symmetric
functions of the square of the particle co-ordinates gives a polynomial
solution. For the sake of illustration, we choose here the power sum basis
$P_l\equiv \sum_i {(x_i^2)}^l$. The eigenfunctions and the energy eigenvalues
are respectively given by
\begin{equation}
\psi_n = \psi_0 \left[\exp\{\hat F/2\} \prod_{l=1}^N P_l^{n_l}\right] \qquad,
\end{equation}
and
\begin{equation}
E_n = 2 \sum_l^N l n_l + E_0 \,; \qquad n = \sum_l l n_l \quad.
\end{equation}
Note that, $\hat S {\hat T}^{-1}$ maps all eigenfunctions
of BnCSM to the even sector of CSM and hence the connection mentioned
earlier between the CSM and BnCSM is valid only in the even sector.

\subsection{Underlying algebraic structure of the BnCSM}

Akin to the CSM case, one can obtain the creation and annihilation
operators for the BnCSM from Eq. (\ref{diabn});
$$b_i^+ = \hat S x_i {\hat S}^{-1} \quad,$$
and
$$b_i^- = \hat  \partial_{x_i} {\hat S}^{-1} \quad,$$
such that, the symmetrized form of the operators,
\begin{eqnarray}
K_i^+ = \frac{1}{2} {b_i^+}^2 \quad,
\end{eqnarray}
acts on the ground-state obtained from
$$b_i^- |0> = 0 \,; \quad i = 1, 2,\cdots, N$$
and creates the eigenstates of the BnCSM. In terms of the creation and
annihilation operators, the BNCSM Hamiltonian can be written as
\begin{eqnarray}
H = \sum_i H_i + (E_0 - N/2) \quad,
\end{eqnarray}
where,
$$H_i \equiv b_i^+ b_i^- + \frac{1}{2} \quad.$$
It can be easily checked that, the commutation relations
\begin{eqnarray}
[K_i^-\,\,,\,\,K_j^+] = \delta_{ij} H_i \quad,
\end{eqnarray}
and
\begin{eqnarray}
[H_i\,\,,\,\,K_j^{\pm}] = \pm 2 \delta_{ij} K_i^{\pm} \quad,
\end{eqnarray}
generate $N$ copies of $SU(1,1)$ algebra; here,
$$K_i^- = \frac{1}{2} {b_i^-}^2 \quad.$$
It is worth noticing that these, $N$, commuting $SU(1,1)$ generators
act on the even sector of the harmonic oscillator basis and hence generate
a complete set of eigenfunctions. Thus, we conclude that, the
underlying algebraic structure of the BnCSM is $SU(1 , 1)$.
In comparision, the eigenfunctions of the CSM contain both the even and
odd sectors of the oscillator basis.\cite{prb}

With the generators of the above mentioned $SU(1,1)$ algebras of BnCSM, one can
define a linear $W_\infty$ algebra for which there exist several basis sets.
\cite{inftyw}

\subsection{New innerproduct}

In parallel to the case of CSM, one can also define
$<<0|{S}_n(\{\frac{1}{2}{b_i^-}^2\}) = <<n|$ and
${S}_n(\{\frac{1}{2}{b_i^+}^2\})|0> = |n>$ as the bra and ket
vectors; ${S}_n$ is a symmetric homogeneous function of degree $n$
and $<<0|\frac{1}{2}{b_i^+}^2 = \frac{1}{2}{b_i^-}^2 |0> = 0$.
Since all the $N$  $SU(1,1)$ algebras are decoupled, the inner
product between these bra and ket vectors proves that any ket
$|n>$, with a given partition of $n$, is orthogonal to all the bra
vectors, with different $n$ and also to those with different
partitions of the same $n$. The normalization for any state $|n>$
can also be found out from the ground state normalization, which
is known. \cite{baker}

\subsection{Quantum integrability}

The quantum integrability of the BnCSM and the identification of the
constants of motion become transparent after establishing its equivalence
to free oscillators. It is easy to verify that $[H\,\,,\,\,H_k] =
[H_i\,\,,\,\,H_j] = 0$ for $i,j,k = 1,2,\cdots, N$. Therefore, the set
$\{H_1, H_2,\cdots, H_N\}$ provides the $N$ conserved quantities. One can
construct linearly independent symmetric conserved quantities analogous to
that of CSM. Here, we would like to point out that, the present proof is
entirely different from earlier works.\cite{bnqi1,bnqi2,bnqi3,bnqi4}

\subsection{The Haschke-R\"uhl model}

Now, we analyse a one-dimensional model of four identical particles, with
both two-body and four-body inverse-square interactions given by the
Hamiltonian,\cite{ruhl}
\begin{equation}
H = - \frac{1}{2} \sum_{i=1}^4 \partial_{x_i}^2 + \frac{1}{2} \sum_{i=1}^4
x_i^2 + g_1 \sum_{{i,j}\atop {i\ne j}} (x_i - x_j)^{-2} + g_2 \sum_{{3 {\rm
independent}}\atop {{\rm terms}}} (x_i + x_j - x_k - x_l)^{-2} \qquad.
\end{equation}
The correlated bosonic ground-state of $H$ is given by
$$\psi_0 = \prod_{i<j} |x_i - x_j|^\alpha
\,\,\,\prod_{\rm 3 indep. terms} (x_i + x_j - x_k - x_l)^\beta
\,\,\,\exp\{- \frac{1}{2} \sum_i x_i^2\}\qquad,$$
where, $\alpha = \frac{1}{2} (1 + \sqrt{1 + 4 g_1})$ and $\beta = \frac{1}{2}
(- 1 + \sqrt{1 + 2 g_2})$. By performing a ST on $H$ with respect to
$\psi_0$, one gets
\begin{equation}
\psi_0^{-1} H \psi_0 \equiv \tilde{H} = \sum_i x_i \partial_{x_i} - \hat{D} +
E_0 \qquad,
\end{equation}
where,
$$\hat{D} = \frac{1}{2} \sum_i \partial_{x_i}^2 + \alpha \sum_{i\ne j}
\frac{1}{x_i - x_j} \partial_{x_i} + \beta \sum_{3 {\rm indep. terms}}
\frac{1}{(x_i + x_j - x_k - x_l)} (\partial_{x_i} + \partial_{x_j} -
\partial_{x_k} - \partial_{x_l})$$
and $E_0 = 2 + 6 \alpha + 3 \beta$. Another ST by $\hat{I} = \exp\{-
\hat{D}/2\}$ on $\tilde{H}$ diagonalizes it completely:
\begin{equation} \label{eul}
\hat{I}^{-1} \tilde{H} \hat{I} \equiv \bar{H} = \sum_i x_i \partial_{x_i} +
E_0 \qquad.
\end{equation}
The explicit connection of $H$ with the decoupled oscillators can be obtained
by one more ST on $\bar{H}$
\begin{equation}
\hat{T}^{-1} \bar{H} \hat{T} = - \frac{1}{2} \sum_{i=1}^4 \partial_{x_i}^2 +
\frac{1}{2} \sum_{i=1}^4 x_i^2 + E_0 - 2 \qquad,
\end{equation}
where, $\hat{T} \equiv \exp(- \frac{1}{2} \sum_{i=1}^4 x_i^2)
\exp(- \frac{1}{4} \sum_{i=1}^4 \partial_{ x_i}^2)$.

Eigenfunctions for this model can be constructed straightforwardly
following the previous examples.

\section{Method for Construction of new solvable models}

More general interacting models of the CSM type, which can be mapped to
decoupled oscillators, can be constructed in the following manner.

One starts with a general Hamiltonian of the type
\begin{equation}\label{gh}
H = - \frac{1}{2} \sum_{i=1}^N \partial_{x_i}^2 + V(x_1,x_2,x_3,\cdots,x_N)
\qquad,
\end{equation}
having $\Phi_0$ as the ground-state wave function and
$$V(x) = \epsilon_0 + \frac{1}{2 \Phi_0} \sum_{i=1}^N \partial_{x_i}^2
\Phi_0 \quad,$$
where, $\epsilon_0$ is the ground-state energy. In order to bring $H$ to
the following form
\begin{equation} \label{gtilde}
\Phi_0^{-1} H \Phi_0 \equiv \tilde{H} = \sum_i^N x_i \partial_{x_i} - \hat{A}
+ \epsilon_0 \qquad,
\end{equation}
the ground-state wave function must be of the form $\Phi_0 = G J$; where,
$$\hat{A} \equiv \frac{1}{2} \sum_{i=1}^N \partial_{x_i}^2 +
\sum_{i=1}^N \partial_{x_i}(\ln J) \partial_{x_i} \quad,$$
and $J$ is so far an arbitrary function.

$\tilde{H}$ can be mapped to the Euler operator by another ST
\begin{equation}
\hat{S}^{-1} \tilde{H} \hat{S} \equiv \bar{H} = \sum_{i=1}^N x_i
\partial_{x_i} + \epsilon_0 \qquad,
\end{equation}
provided, the following equation holds,
\begin{equation}
\left[\tilde H , \exp\{- \hat A/2\}\right] =
\left[\sum_i x_i {\partial}_{x_i} , \exp\{- \hat A/2\}\right]
= \hat A \exp\{- \hat A/2\} \qquad.
\end{equation}
The above condition restricts $J$ to be a homogeneous function of the
particle coordinates.

Now, it is easy to see that, the Hamiltonian in Eq.(\ref{gh}) can be mapped to
free oscillators by a series of STs,
\begin{equation}
G \hat{E}(0) \exp\{\hat A/2\} {\Phi_0}^{-1} H \Phi_0 \exp\{- \hat A/2\} {\hat
E}^{-1}(0) G^{-1} = -\frac{1}{2} \sum_i \partial_{x_i}^2 +
\frac{1}{2} \sum_i x_i^2 + (\epsilon_0 - \frac{1}{2} N) \,.
\end{equation}

However, as mentioned earlier, it is important to check that, the action
of $\exp\{- \hat A/2\}$ on
an appropriate linear combination of the eigenstates of $\sum_{i=1}^N x_i
\partial_{x_i}$, yields solutions which are normalizable with respect to
$\Phi_0^2$ as the weight function. Appropriate choices of $J$ will yield
new solvable models having linear spectra. This can also be
generalized to the higher-dimensional interacting models,
\cite{guru1,pkg,srg} provided one is careful about the intricacies arising
in higher dimensions.

\section{A New technique to solve linear differential equations and
diagonalization of the Sutherland model}

In the previous section, we have dealt with a class of many-body
Hamiltonians, which, after a similarity transformation separate
into a part containing the Euler operator and a constant and
another operator $\hat{A}$ having a definite degree. These
Hamiltonians were then diagonalized, taking advantage of the fact
that $[\sum_{i}x_i
\partial_{x_i}, e^{\hat{A}/d}] = \hat{A} e^{\hat{A}/d}$, where d is the degree
of the operator $\hat{A}$, i.e., $[\sum_{i}x_i \partial_{x_i},
\hat{A}] = d \hat{A}$. We notice that the above procedure fails in
the cases, where the similarity transformed Hamiltonian
$\tilde{H}$ contains an operator $\hat{F}$ possessing the
following properties.

\noindent {\it Case i}: When $\hat F = \sum_{k=1}^s \hat F_k$,
where, $s$ is an arbitrary integer and $\hat F_k$'s are operators
with different degrees and $[\hat F_k \,\,,\,\,F_l] \ne 0$; for
$k, l = 1, 2, \cdots s$.

\noindent {\it Case ii}: When the degree, $d$, of the opertor
$\hat F$ is zero.

\noindent {\it Case iii}: When $\tilde H$ contains operators like
$\sum_i (x_i \partial_{x_i})^n $, with $n \ge 2$.

As will be shown below, one needs a method which works for the
second and third cases mentioned above, in order to diagonalize
the Sutherland model and to show its equivalence to free particles
on a circle.

A recently proposed method \cite{cha1} is illustrated below
which overcomes all the difficulties mentioned above and achieves
the goal of connecting the solutions of differential equations to
monomials.

Consider first the general, single variable linear differential
equation,
\begin{eqnarray} \label{ie}
\left(F(D) + \hat{P} \right) y(x) = 0 \quad,
\end{eqnarray}
where, $D \equiv x \frac{d}{dx}$ and $F(D) = \sum_{n = -
\infty}^{n = \infty} a_n D^n $, is a diagonal operator in the
space of monomials. $\hat{P}$ can be an arbitrary operator, having
a well-defined action in the space spanned by $x^n$. Here, $a_n$'s
are some parameters.

The solution of Eq. (\ref{ie}) is given by
\begin{eqnarray} \label{an}
y(x) &=& C_\lambda \left \{\sum_{m = 0}^{\infty} (-1)^m
\left[\frac{1}{F(D)} \hat{P} \right]^m \right \} x^\lambda \nonumber\\
&\equiv& C_\lambda \hat{G}_\lambda x^\lambda \qquad,
\end{eqnarray}
provided, $F(D) x^\lambda = 0$ and the coefficient of $x^\lambda$
in $y(x) - C_\lambda x^\lambda$ is zero (no summation over
$\lambda$); here, $C_\lambda$ is a constant. The latter condition
guarantees that, the solutions, $y(x)$'s, are non-singular. The proof is
straightforward and follows by direct substitution. \cite{cha1,cha2}
Note that, the detailed properties of $\hat{P}$ are not needed to
prove that $y(x)$ in Eq. (\ref{an}) is a solution of Eq.
(\ref{ie}). However, naturally, these are required while
constructing the explicit solutions of any given linear
differential equation. This procedure gives novel expression for
the solutions of known differential equations and brings out the
properties of the orthogonal polynomials and functions in a
natural way.\cite{cha1,cha2}

Eq. (\ref{an}), which connects the solutions of a differential
equation to the monomials, can be generalized to many-variables as
follows.

Consider,
\begin{eqnarray} \label{soli}
\left(\sum_{n = -\infty}^\infty b_n (\sum_i D_i^n) + \hat{A}
\right) Q_\lambda(x) = B_\lambda(x) \qquad,
\end{eqnarray}
where, $b_n$'s are some parameters, $D_i \equiv x_i
\partial_{x_i}$; $\hat{A}$ can be a function of $x_i$,
$\partial_{x_i}$ and also some other well-defined composite operators 
and $B_\lambda(x)$ is a source
term. Solutions of Eq. (\ref{soli}) can be obtained for different
cases.

\noindent {\it Case (i)}: When $B_\lambda(x) = 0$ and $\hat{A}
m_\lambda = \epsilon_\lambda m_\lambda + \sum_{\mu < \lambda}
C_{\mu \lambda} m_{\mu}$; where, $m_\lambda$'s are the monomial
symmetric functions \cite{jack4} and $\epsilon_\lambda$ and
$C_{\lambda \mu}$ are some coefficients.

Using Eq. (\ref{an}), the solution can be obtained as,
\begin{eqnarray} \label{4sm5}
Q_\lambda(x) = \sum_{r=0}^\infty (- 1)^r \left[\frac{1}{((\sum_{n
= -\infty}^\infty b_n (\sum_i D_i^n) - (\sum_{n = -\infty}^\infty
b_n (\sum_i {\lambda}_i^n))} (\hat{A} - \epsilon_\lambda)\right]^r
m_\lambda(x) \,\,
\end{eqnarray}
with, $\sum_{n = -\infty}^\infty b_n (\sum_i {\lambda}_i^n) +
\epsilon_ \lambda = 0$.

\noindent {\it Case (ii)}: When $B_\lambda(x) \ne 0$.

\begin{eqnarray}
Q_\lambda(x) = \sum_{r=0}^\infty (- 1)^r \left[\frac{1}{((\sum_{n
= -\infty}^\infty b_n (\sum_i D_i^n) - (\sum_{n = -\infty}^\infty
b_n
(\sum_i {\lambda}_i^n))} \hat{A}\right]^r \nonumber\\
\times \left[\frac{1}{((\sum_{n = -\infty}^\infty b_n (\sum_i
D_i^n) - (\sum_{n = -\infty}^\infty b_n (\sum_i
{\lambda}_i^n))}\right] B_\lambda(x) \,\,,
\end{eqnarray}
provided, the coefficient of the divergent part in the right hand
side of the above equation is zero. As will be soon seen in the
context of the SM, this requirement yields the eigenvalues for the
Hamiltonians, with which the above differential equations are
associated.

We now proceed to use the above mentioned method to diagonalize
the SM and establish a precise connection between the Hilbert
space of SM and that of free particles on a circle. The
Schr\"odinger equation is,
\begin{eqnarray} \label{sut1}
\left(- \sum_{i=1}^N \frac{\partial^2}{ \partial x_i^2} + 2 \beta
(\beta - 1) \frac{\pi^2}{L^2} \sum_{i<j} \frac{1}{\sin^2[\pi (x_i
- x_j)/L]} - E_\lambda \right) \psi_\lambda(\{x_i\}) = 0 \qquad.
\end{eqnarray}
Choosing, $z_j = e^{2\pi i x_j /L}$ and writing
$\psi_\lambda(\{z_i\}) = \prod_i z_i^{- (N - 1) \beta /2}
\prod_{i<j} (z_i - z_j)^\beta J_\lambda(\{z_i\})$, the above
equation becomes,
\begin{eqnarray} \label{jac}
\left(\sum_i D_i^2 + \beta \sum_{i<j} \frac{z_i + z_j}{z_i - z_j}
(D_i - D_j) + \tilde{E}_0 - \tilde{E}_\lambda \right)
J_\lambda(\{z_i\}) = 0 \qquad,
\end{eqnarray}
where, $D_i \equiv z_i \frac{\partial}{\partial z_i}$,
$\tilde{E_\lambda} \equiv (\frac{L}{2 \pi})^2 E_\lambda$,
$\tilde{E_0} \equiv (\frac{L}{2 \pi})^2 E_0$ and $E_0 =
\frac{1}{3}(\frac{\pi}{L})^2 \beta^2 N (N^2 - 1)$, is the
ground-state energy. Here, $J_\lambda(\{z_i\})$'s are known as the
Jack polynomials.\cite{jack4,jack1,jack2,jack3} $\sum_i D_i^2$ is
a diagonal operator in the space spanned by the monomial symmetric
functions, $m_{\{\lambda\}}$, with eigenvalues $\sum_{i=1}^N
\lambda_i^2$. Rewriting Eq. (\ref{jac}) in the form,
\begin{eqnarray}
\left(\sum_i (D_i^2 - \lambda_i^2) + \beta \sum_{i<j} \frac{z_i +
z_j}{z_i - z_j} (D_i - D_j ) + \tilde{E}_0 + \sum_i \lambda_i^2 -
\tilde{E}_\lambda \right) J_\lambda(\{z_i\}) = 0 \qquad,
\end{eqnarray}
and using Eq. (\ref{4sm5}), one can immediately show that,
\begin{eqnarray} \label{sr}
J_\lambda(\{z_i \}) &=& C_\lambda \left \{\sum_{n = 0}^{\infty}
(-1)^n \left[\frac{1}{\sum_i (D_i^2 - \lambda_i^2)}(\beta
\sum_{i<j} \frac{z_i + z_j}{z_i - z_j}(D_i - D_j ) +  \tilde{E}_0
+ \sum_i \lambda_i^2 -
\tilde{E}_\lambda)\right]^n \right \} \nonumber\\
&& \qquad \qquad \qquad \qquad \qquad \qquad \qquad \qquad \qquad
\qquad \qquad \qquad \qquad \times  m_\lambda(\{z_i\}) \nonumber\\
&\equiv& C_\lambda \hat{G}_\lambda m_\lambda(\{z_i \}) \qquad.
\end{eqnarray}
It can be checked that, $\hat{G}_\lambda$ maps the SM to free
particles on a circle, {\it i.e.},
\begin{eqnarray}
(\psi_0 \hat{G}_\lambda)^{-1} H_S (\psi_0 \hat{G}_\lambda) &=&
\left(\frac{2 \pi}{L}\right)^2 \left(\sum_i D_i^2 - \sum_i
\lambda_i^2 +
\tilde{E}_\lambda \right) \nonumber\\
&=& - \sum_{i=1}^N \frac{\partial^2}{ \partial x_i^2} -
\left(\frac{2 \pi}{L}\right)^2 \sum_i \lambda_i^2 + {E}_\lambda
\qquad,
\end{eqnarray}
where, $H_S$ is the SM Hamiltonian and $\psi_0$ is its
ground-state wave function. For the sake of convenience, we define
\begin{eqnarray} \label{S}
\hat{S} & \equiv & \left[\frac{1}{\sum_i (D_i^2 - \lambda_i^2)}
\hat{Z} \right] \qquad, \nonumber\\
\mbox{and} \qquad \hat{Z} & \equiv & \beta \sum_{i<j} \frac{z_i +
z_j}{z_i - z_j} (D_i - D_j) +  \tilde{E}_0 + \sum_i \lambda_i^2 -
\tilde{E}_\lambda \qquad.
\end{eqnarray}
The action of $\hat{S}$ on $m_\lambda(\{z_i \})$ yields
singularities, unless one chooses the coefficient of $m_\lambda$
in $\hat{Z} \, m_\lambda(\{z_i\})$ to be zero; this condition
yields the well-known eigenspectrum of the Sutherland model:
$$
\tilde{E}_\lambda = \tilde{E}_0 + \sum_i (\lambda_i^2 + \beta [N +
1 - 2 i] \lambda_i) \qquad.
$$
Using the above, one can write down a novel expression for the
Jack polynomials as,
\begin{eqnarray} \label{nfj}
J_\lambda(\{z_i \}) = \sum_{n=0}^\infty (- \beta)^n
\left[\frac{1}{\sum_i (D_i^2 - \lambda_i^2)} (\sum_{i<j} \frac{z_i
+ z_j}{z_i - z_j}(D_i - D_j) - \sum_i
(N + 1 - 2 i) \lambda_i )\right]^n \nonumber\\
\qquad \qquad \qquad \qquad \qquad  \times m_\lambda(\{x_i\})
\qquad.
\end{eqnarray}
 This procedure for constructing the Jack polynomials does not involve
 additional mathematical tools like Dunkl derivatives and $S_{N}$ extended
 Heisenberg algebra.\cite{psu2} The implications of the just established connection between
 SM wave functions and free particles are currently under
 investigation. In particular, its relevance to generalized
 statistics is of deep interest.\cite{apo} The connection of Jack
 polynomials with the monomial symmetric functions can be used for
 constructing ladder operators, since at the level of the monomials, the construction
 of the ladder operators is straightforward.\cite{cha2}
 Currently, these operators have been constructed only for a few
 particle system.\cite{gar} We are also studying the symmetry
 properties of the Jack and other multivariate polynomials, using the above
 technique. It is worth pointing out that Jack polynomial and its supersymmetric
 generalizations are being studied extensively, for their relevance to multi-variate
 statistics and mathematical physics.\cite{sups1,sups} 
 It is easy to see that, the method
 presented above generalizes to other Sutherland type models. The
 straightforwardness of this approach should be contrasted with an earlier
 approach to these models by the present authors, through $S_N$ extended
 Heisenberg algebra and operators related to them.\cite{gurappa}

\section{Laughlin wave function and decoupled harmonic oscillators}

In this section, we study a planar many-body Hamiltonian relevant for
the description of the quantum Hall effect.\cite{qhe} This Hamiltonian
describes
electrons in a magnetic field, with two-body and three-body inverse-square
interactions arising due to the Chern-Simons gauge field and have Laughlin
wave function\cite{lin} as the ground-state.\cite{srg} We explicitly
prove that, these models can be exactly mapped to a set of free harmonic
oscillators on the plane. As a consequence, the existence of {\it
linear} $W_{1+\infty}$ algebra with Laughlin wave function as its highest
weight vector, is pointed out, in a rather elegant and straightforward
manner. This study also brings out the potential difficulties one can face
in more than one dimension.

We start with the Hamiltonian:
\begin{eqnarray} \label{iz}
H =&& \frac{1}{2} \sum_{i=1}^N ( - 4 \partial_{{\bar{z}}_i} \partial_{z_i} +
z_i \partial_{z_i} - \bar{z}_i \partial_{\bar{z}_i} + \frac{1}{4} \bar{z}_i
z_i ) + 2 \eta \sum_{{i=1\atop {i\ne j}}}^N \frac{1}{(z_i - z_j)}
(\partial_{\bar{z}_i} - \frac{1}{4} z_i) \nonumber\\
&-& 2 \eta \sum_{{i=1\atop {i\ne j}}}^N \frac{1}{(\bar{z_i} - \bar{z_j})}
(\partial_{z_i} - \frac{1}{4} \bar{z_i})
+ 2 \eta^2 \sum_{{i,j,k=1}\atop
{i\ne j ; i\ne k}}^N \frac{1}{(z_i - z_j) (\bar{z}_i - \bar{z}_k)}
\quad.
\end{eqnarray}
The ground-state of this model was found to be of Laughlin form,
$$\psi_0 =
\prod_{i<j} (z_i - z_j)^\eta \,\,\exp\{- \frac{1}{4} \sum_i \bar{z}_i
z_i\}\qquad.$$
By performing a ST, one gets
\begin{eqnarray} \label{st}
\psi_0^{-1} H \psi_0 \equiv \tilde{H} = \sum_i z_i \partial_{z_i} - \hat{A}
+ \frac{1}{2} N \qquad,
\end{eqnarray}
where,
$$\hat{A} \equiv 2 \sum_i \partial_{{\bar{z}}_i}
\partial_{z_i} + 2 \eta \sum_{i\ne j} (\frac{1}{\bar{z_i} - \bar{z_j}}
\partial_{z_i}- \frac{1}{z_i - z_j}
\partial_{{\bar{z}}_i})\quad.$$
It is easy to check that
\begin{equation}
[\sum_i z_i \partial_{z_i}\,\,,\,\,\hat{A}] = - \hat{A} + 4 \pi \eta
\sum_{{i,j}\atop {i\ne j}} (z_i - z_j) \delta^2(z_i - z_j) \partial_{z_i}
\qquad.
\end{equation}
In view of the identity {\it i.e.}, $x \delta(x) = 0$,
the above equation reduces to
\begin{equation}
[\sum_i z_i \partial_{z_i}\,\,,\,\,\hat{A}] = - \hat{A} \qquad.
\end{equation}
Performing a ST by $\exp\{- \hat{A}\}$, Eq. (\ref{st}) becomes
\begin{eqnarray}\label{fz}
\exp\{\hat{A}\}\,\,\tilde{H}\,\,\exp\{- \hat{A}\} \equiv \bar{H} = \sum_i
z_i \partial_{z_i} + \frac{1}{2} N \qquad.
\end{eqnarray}
Finally, the following ST by $\hat{W} \equiv \exp\{2 \sum_i
\partial_{{\bar{z}}_i} \partial_{z_i}\} \exp\{\frac{1}{4} \sum_i \bar{z}_i
z_i\}$ brings the above Hamiltonian to a Hamiltonian of $N$ free
harmonic oscillators,
\begin{equation}
\hat{W}^{-1} H \hat{W} = \frac{1}{2} \sum_{i=1}^N ( - 4
\partial_{{\bar{z}}_i} \partial_{z_i} + z_i \partial_{z_i} - \bar{z}_i
\partial_{\bar{z}_i} + \frac{1}{4} \bar{z}_i z_i) \qquad.
\end{equation}
By defining $a_i^+ = {\hat{S}}^{-1} z_i \hat{S}$ and $a_i^- = {\hat{S}}^{-1}
\partial_{z_i} \hat{S}$; where $\hat{S} = \psi_0 \exp\{\hat{A}\}$, and making
use of Eq. (\ref{fz}), one can rewrite Eq. (\ref{iz}) as
\begin{equation}
H = \sum_i H_i + \frac{1}{2} N = \sum_i a_i^+ a_i^- + \frac{1}{2} N
\end{equation}
where, $H_i \equiv a_i^+ a_i^-$, such that
$$[a_i^-\,\,,\,\,a_j^+] = \delta_{ij}\qquad,$$
and
$$[H_i\,\,,\,\,a_j^{\pm}] = \pm a_j^\pm \delta_{ij}\qquad.$$
These $N$ quantities $H_i$ serve as the conserved quantities and are in
involution, {\it i.e.}, $[H_i\,\,,\,\,H_j] = 0$.

Since $a_i^-$ and $a_i^+$ obey non-interacting oscillator algebra, one can
make use of this fact to define a {\it linear} $W_{1+\infty}$ algebra,
making use of the basis given in Ref. [\onlinecite{zemb}].
The highest weight vector obtained from $L_{m,n} \psi_0 = 0$ for $n > m
\ge -1$ is nothing but the Laughlin wave function.

\section{Conclusions and Discussions}

In conclusion, we have extended the procedure for 
the exact diagonalization of the N-particle
Calogero-Sutherland model to a host of related models and have 
demonstrated the precise
correspondence between these models and free oscillators. The underlying
spectrum generating algebras, the construction of the excited
state eigenfunctions and the conserved quantities responsible for
the quantum integrability were also explicated. Although, we have
worked out only the $A_{N-1}$ and $B_N$ type models, for the sake
of brevity, it is evident that our method generalizes to other
root systems. In light of the fact that, the eigenvectors of the
CSM have correspondence with the singular vectors of the $W_N$
algebras,\cite{wn} it is worth studying in greater detail, the
properties of our exact expressions for the excited state
eigenvectors. This will throw more light on the emergence of
conformal field theory results\cite{cft1,cft2,cft3} from the CSM
in various limiting conditions. The construction of the coherent
states for these correlated systems is now straightforward,
starting from the known coherent states of the oscillator systems.
In the two particle case, it is known that, the interaction brings
in  interesting behaviour in the coherent states.\cite{gsasc}
Hence, the N-particle case needs further investigation. The method
elaborated here can also be applied to other models involving
nearest and next-to-nearest neighbour
interactions.\cite{nnn1,nnn2,nnn3,nnn4} These models are related
to pseudo-integrable systems, which are currently attracting
considerable attention in connection with physical systems like
metal-insulator transitions.\cite{metin} The procedure for
diagonalizing the CSM has been found inadequate for dealing with
SM and its generalizations. A recently developed method was used
for mapping the SM to free particles on a circle, which also
yielded a novel connection between the Jack polynomial and
monomial symmetric functions. The precise relationship between the
wave functions of SM and the free particle eigenfunctions is
currently under study, in light of its relevance to generalized
statistics as also the problem of finding ladder operators for the
SM.

In two dimensions, the consequences of the linear $W_{1+\infty}$
symmetry algebra, the realization of the Laughlin wave function as
the highest weight vector and the oscillator connection of the
Chern-Simons Hamiltonians need further exhaustive study. Apart
from the deep interest of the lowest Landau level physics, the
aforementioned Hamiltonians involving Chern-Simons gauge fields,
also provide realizations of the Jain picture.\cite{jain} It is
worth mentioning that, the lowest Landau level physics can be
mapped to a one dimensional quantum mechanical system of the
Calogero-Sutherland type.\cite{qhe1,qhe2,qhe3} Hence, the results
of this paper for one dimensional models may find relevance for
electrons in a strong magnetic field. Further investigations,
along the above lines are currently under progress and will be
reported elsewhere. Recently, Calogero-Sutherland type models are
also being extended to non-commutative spaces,\cite{apo1} which
has relevance to lowest Landau level physics.\cite{biga,suss}
Hence, extending the present approach to the noncommutative space
is also an interesting area to explore.


\begin{thebibliography}{99}

\bibitem{cal1} F. Calogero, {\it Jour. Math. Phys.} {\bf 12}, 419 (1971).

\bibitem{suther1} B. Sutherland, {\it Jour. Math. Phys.} {\bf 12}, 246 (1971).

\bibitem{suther2} B. Sutherland, {\it Jour. Math. Phys.} {\bf 12},
251 (1971).

\bibitem{suthd} B. Sutherland, {Phys. Rev.} {\bf A 4}, 2019
(1971); {\it ibid.} {\bf 5}, 1372 (1972).

\bibitem{ms1} S. Tewari, {\it Phys. Rev.} {\bf B 46}, 7782 (1992).

\bibitem{ms2} N.F. Johnson and M.C. Payne, {\it Phys. Rev. Lett.} {\bf
70}, 1513 (1993); {\it ibid.} {\bf 70}, 3523 (1993).

\bibitem{ms3} M. Caselle, {\it Phys. Rev. Lett.} {\bf 74}, 2776 (1995).

\bibitem{qhe1} N. Kawakami, {\it Phys. Rev. Lett.} {\bf 71}, 275 (1993).

\bibitem{qhe2} H. Azuma and S. Iso, {\it Phys. Lett.} {\bf B 331}, 107
(1994).

\bibitem{qhe3} P.K. Panigrahi and M. Sivakumar, {\it Phys. Rev.} {\bf B
52}, 13742 (15) (1995).

\bibitem{wp} H.H. Chen, Y.C. Lee and N.R. Pereira, {\it Phys. Fluids} {\bf
22}, 187 (1979).

\bibitem{rmt1} F.J. Dyson, {\it J. Math. Phys.} {\bf 3}, 140, 157, 166
(1962).

\bibitem{rmt2} F.J. Dyson, {\it J. Math. Phys.} {\bf 19}, 235 (1970).

\bibitem{rmt3} B.L. Altshuler, I.K. Zharekeshev, S.A. Kotochigova and B.I.
Shklovskii, {\it Sov. Phys. JETP.} {\bf 67}, 625 (1988).

\bibitem{rmt4} M.L. Mehta, {\it Random Matrices}, Revised Edition
Academic Press, New York, 1990.

\bibitem{rmt5} M. Ndawana, R.A. Romer and M. Schreiber, {\it Eur. Phys. J.} {\bf
B 27}, 399 (2002).

\bibitem{nala} K. Nakamura and M. Lakshmanan, {\it Phys. Rev. Lett.}
{\bf 57,} 1661 (1986).

\bibitem{sim} B. D. Simons, P. A. Lee and B. L. Altshuler, {\it Phys. Rev. Lett.}
{\bf 72,} 64 (1994) and references therein.


\bibitem{fs1} J.M. Leinaas and J. Myrheim, {\it Phys. Rev.} {\bf B 37},
9286 (1988).

\bibitem{fs2} F.D.M. Haldane, {\it Phys. Rev. Lett.} {\bf 67}, 937 (1988).

\bibitem{fs3} A.P. Polychronakos, {\it Nucl. Phys.} {\bf B 324}, 597
(1989).

\bibitem{fs4} M.V.N. Murthy and R. Shankar, {\it Phys. Rev. Lett.} {\bf
73}, 3331 (1994).

\bibitem{fs5} Z.N.C. Ha, {\it Phys. Rev. Lett.} {\bf 73}, 1574 (1994).

\bibitem{gra1} I. Andric, A. Jevicki and H. Levine, {\it Nucl. Phys.}
{\bf B 312}, 307 (1983).

\bibitem{gra2} A. Jevicki, {\it Nucl. Phys.} {\bf B 376}, 75 (1992).

\bibitem{gra3} G.W. Gibbons and P.K. Townsend, {\it Phys. Lett.}
{\bf B 454}, 187 (1999).

\bibitem{anyon1} A. Lerda, {\it Anyons}, Lecture notes in physics, Eds.
H. Araki {\it et al.}, Springer-Verlag, Berlin Heidelberg, 1992
and references therein.

\bibitem{anyon2} S.B. Fsakov, G. Lozano and S. Ouvry, {\it Non-Abelian
Chern-Simons Particles in an External Magnetic Field},
hep-th/9902028.

\bibitem{anyon3} S. Ouvry, {\it On the relation between anyon and
Calogero models}, cond-mat/9907239.

\bibitem{gt1} J.A. Minahan and A.P. Polychronakos, {\it Phys. Lett.}
{\bf B 312}, 155 (1993); {\it ibid.} {\bf 336}, 288 (1994).

\bibitem{gt2} E. D'Hoker and D.H. Phong, {\it Nucl. Phys.} {\bf B 513}, 405
(1998).

\bibitem{book} {\it Calogero-Moser-Sutherland models}, Eds. J. F. Van Diejen and L.
Vinet, CRM series in Mathematical Physics, Springer, 2000.

\bibitem{moap} M. Olshanetsky and A. Perelomov, {\it Phys. Rep.}
{\bf 71}, 314 (1981); {\it ibid.} {\bf 94}, 6 (1983).

\bibitem{apo} A.P. Polychronakos, {\it 'Generalized statistics in one
dimension,'} published in {\it 'Topological aspects of
low-dimensional systems,'} Les Houches session LXIX (1998),
Springer Ed., hep-th/9902157.

\bibitem{adh} E. D'Hoker and D. H. Phong, hep-th/9912271.

\bibitem{ajb} A. J. Bordner, E. Corrigan and R. Sasaki, {\it Prog.
Theor. Phys.} {\bf 102}, 499 (1999).

\bibitem{pi} J. Dukelsky, C. Esebbag and P. Schuck, {\it Phys. Rev. Lett.} {\bf
87}, 066403-1 (2001).

\bibitem{pt} F. Iachello, {\it Phys. Rev. Lett.} {\bf85}, 3580
(2000).

\bibitem{gra4} S. Corley, {\it JHEP}, {\bf 9909}, 187 (1999); D. Birmingham, K.
S. Gupta and S. Sen, {\it Phys. lett.} {\bf B 505}, 191 (2001).

\bibitem{exop} M. Vasiliev, {\it Int. J. Mod. Phys.} {\bf A 6}, 1115
(1991).

\bibitem{eop} A.P. Polychronakos, {\it Phys. Rev. Lett.} {\bf 69}, 703
(1992).

\bibitem{yan1} K. Hikami, {\it J. Phys.} {\bf A}: {\it Math. Gen.} {\bf
28}, L131 (1995).

\bibitem{yan2} V.B. Kuznetsov, {\it Phys. Lett.} {\bf A 218}, 212 (1996).

\bibitem{dea} J. de Boer, F. Harmsze and T. Tjin, {\it Phys. Rep.} {\bf
272}, 139 (1996) and references therein.

\bibitem{vel1} V. Bardek, L. Jonke, S. Meljanac and
M. Melekovic, {\it Calogero model, deformed oscillators and the
collapse}, hep-th/0107053, to appear in Phys. Lett. {\bf B}.

\bibitem{vel2} V. Bardek, S. Meljanac, {\it Eur. Phys.J.} {\bf C
17}, 539 (2000).

\bibitem{cal2} F. Calogero, {\it J. Math. Phys.} {\bf 10}, 2191 (1969).

\bibitem{prb} N. Gurappa and P.K. Panigrahi, {\it Phys. Rev.} {\bf B},
R2490 (1999).

\bibitem{uji1} H. Ujino, A. Nishino and M. Wadati, {\it J. Phys. Soc.
Jpn.} {\bf 67}, 2658 (1998); {\it Phys. Lett.} {\bf A 249}, 459 (1998).

\bibitem{sscsm1} P.K. Ghosh, {\it Nucl. Phys.} {\bf B 595}, 519 (2001).

\bibitem{sscsm2} A. Punnoose and R.A. Romer, {\it J. Phys.} {\bf A 29}, 1651
(1996) and references therein.

\bibitem{gon} T. Brzezinski, C. Gonera and P. Maslanka, {\it
Phys.Lett.} {\bf A 254}, 185 (1999).

\bibitem{susy1} B. Basu-Mallik, H. Ujino and M. Wadati, {\it J. Phys. Soc.
Jpn.} {\bf 68}, 3219 (1999).

\bibitem{susy2} A.J. Bordner, N.S. Manton and R. Sasaki, {\it
Calogero-Moser Models V: Supersymmetry and Quantum Lax Pair},
hep-th/9910033.

\bibitem{susy3} T.D. Deguchi and P.K. Ghosh, {\it Spin Chains from Super
Models}, hep-th/0012058.

\bibitem{nnn1} S.R. Jain and A. Khare, {\it Phys. Lett.} {\bf A 262}, 35
(1999).

\bibitem{nnn2} G. Auberson, S.R. Jain and A. Khare, {\it Off-diagonal
long-range order in one-dimensional many-body problem}, cond-mat/9912445,
to appear in {\it Phys. Lett.} {\bf A}.

\bibitem{nnn3} G. Auberson, S.R. Jain and A. Khare, {\it J. Phys.}
{\bf A} (in press), cond-mat/0004012.

\bibitem{nnn4} M. Ezung, N. Gurappa, A. Khare and P.K. Panigrahi, {\it
Algebraic study of a quantum many-body system with nearest neighbour
long-range interactions}, cond-mat/0007005.

\bibitem{bbm} B. Basu-Mallick and A. Kundu, {\it Phys. Rev.} {\bf B 62}, 9927 (2000).

\bibitem{cha1} N. Gurappa, P. K. Panigrahi, T. Shreecharan and
S. Sree Ranjani, {\it Frontiers of fundamental physics} {\bf 4},
Eds. Sidharth and Altaisky, Kluwer Academic Plenum Publishers, New
York, 2001.

\bibitem{cha2} N. Gurappa, P. K. Panigrahi and T. Shreecharan,
{\it Linear differential equations and orthogonal polynomials: A
novel approach}, math-ph/0203015.

\bibitem{ruhl} O. Haschke and W. R\"uhl, pre-print no. KL-TH 98/9,
hep-th/9807194.

\bibitem{qhe} M. Stone, {\it Quantum Hall Effect}, World Scientific,
1992 and references therein.

\bibitem{jain} J.K. Jain, {\it Phys. Rev. Lett.} {\bf 63}, 199 (1984); V.J.
Goldman, B. Su and J.K. Jain, {\it ibid} {\bf 72}, 2065 (1994).

\bibitem{sogo} K. Sogo, {\it J. Phys. Soc. Jpn.} {\bf 65}, 3097 (1996).

\bibitem{jack4} I.G. Macdonald, {\it Symmetric Functions and Hall
Polynomials}, 2nd edition, Oxford: Clarendon press, 1995.

\bibitem{mvlp} A.T. James in {\it Theory and Applications of Special
Functions}, R.A. Askey ed., Academic Press, New York, 1975; R.J.
Muirhead, {\it Aspects of multivariate statistical thory}, John Wiley,
New York, 1982.

\bibitem{inftyw}  S.R. Das, A. Dhar, G. Mandal and S.R. Wadia, {\it Int.
J. Mod. Phys}. {\bf A 7}, 5165 (1992); A. Cappelli, C.A. Trugenberger and
G.R. Zemba, {\it Nucl. Phys}. {\bf B 396}, 465 (1993).

\bibitem{baker} T.H. Baker and P.J. Forrester, solv-int/9608004, 9609010;
{\it Nucl. Phys.} {\bf B 492}, 682 (1997).


\bibitem{bnqi1} M.A. Olshanetsky and A.M. Perelomov, {\it Lett. Math. Phys}.
{\bf 2}, 7 (1977).

\bibitem{bnqi2} I. Cherednik, {\it Adv. Math}. {\bf 106}, 65 (1994).

\bibitem{bnqi3} T. Yamamoto, {\it J. Phys. Soc. Jpn}. {\bf 63}, 1212
(1994).

\bibitem{bnqi4} J.F. van Diejen, {\it Compos. Math}. {\bf 95}, 183 (1995).


\bibitem{guru1} N. Gurappa, A. Khare and P.K. Panigrahi, {\it Phys. Lett.}
{\bf A 224}, 467 (1998).

\bibitem{pkg} P.K. Ghosh, {\it Phys. Lett.} {\bf A 229}, 203 (1997).

\bibitem{srg} R.K. Ghosh and S. Rao, {\it Phys. Lett.} {\bf A 238}, 213
(1998).

\bibitem{jack1} H. Jack, {\it Proc. R. Soc. Edinburgh} (A), {\bf 69}, 1
(1970); 347, (1972).

\bibitem{jack2} R.P. Stanley, {\it Adv. Math.} {\bf 77}, 76 (1988).

\bibitem{jack3} S. Chaturvedi, {\it Special Functions and Differential
Equations}, Eds. K.S. Rao, R. Jagannathan, G.V. Berghe and J.V. Jeugt,
Allied publishers, New Delhi, 1997.

\bibitem{psu2} L. Lapointe and L. Vinet, {\it Commun. Math. Phys.} {\bf
178}, 425 (1996).

\bibitem{gar} W. Garcia Fuertes, M. Lorente and A. M. Perelomov,
{J. Phys.} {\bf A:} {\it Math. Gen.} {\bf 34}, 10963 (2001).

\bibitem{sups1} A. Veselov, M. Feigin and O. Chalykh, 
{\it Comm. Math. Phys.} {\bf 206}, 533
(1999).

\bibitem{sups} A. Sergeev, {\it J. Nonlin. Math. Phys.}, {\bf 8}, 59 (2001).

\bibitem{gurappa} N. Gurappa and P.K. Panigrahi, {\it Phys. Rev.} {\bf B 62},
1943 (2000).

\bibitem{lin} R.B. Laughlin, {\it Phys. Rev. Lett.} {\bf 50}, 1395 (1983).

\bibitem{zemb} A. Cappelli, C.A. Trugenberger and G.R. Zemba, Nucl.
Phys. {\bf B 396}, 465 (1993); {\it Phys. Rev.} {\bf B 306}, 100 (1993).


\bibitem{wn} H. Awata, Y. Matsuo, S. Odke and J. Shiraishi
{\it Phys. Lett.} {\bf B 347}, 49 (1994); {\it Nucl. Phys.} {\bf B 449},
347 (1995).

\bibitem{cft1} N. Kawakami and S.-K. Yang, {\it Phys. Rev. Lett.} {\bf
67}, 2493 (1991).

\bibitem{cft2} M. Fran,  S. Sciuto, A. Lerda and G.R. Zemba, {\it Int. J.
Mod. Phys.} {\bf A 12}, 4611 (1997).

\bibitem{cft3} M. Cadoni, P. Carta and D. Klemm, {\it Phys. Lett.} {\bf B
503}, 205 (2001).

\bibitem{gsasc} G.S. Agarwal and S. Chaturvedi, {\it J. Phys.} {\bf A 28},
5747 (1995).

\bibitem{metin} T. Guhr, A. M\"{u}ller-Groeling and H.A. Weidenm\"{u}ller,
{\it Phys. Rep.} {\bf 299}, 189 (1998).

\bibitem{apo1} A. P. Polychronakos, {\it Phys. Rev. Lett.} {\bf 89}, 126403
(2002).

\bibitem{biga} D. Bigatt and L. Susskind, {\it Phys. Rev.} {\bf D 62}, 066004 (2000).

\bibitem{suss} L. Susskind, {\it The Quantum Hall fluid and noncommutative
Chern-Simons theory}, hep-th/0101029.

\end{thebibliography}
\end{document}